\documentstyle[preprint,aps]{revtex}

\begin{document}
\draft
\preprint{CMU-HEP93-28; DOE-ER/40682-53}
\title{The Origin and Mechanisms of CP Violation In the   \\
  Two-Higgs Doublet Model and Masses of the Exotic Scalars}
\author{Yue-Liang  Wu\cite{byline}}
\address{ Department of Physics, \ Carnegie Mellon University \\ Pittsburgh,
 Pennsylvania 15213,\ U.S.A.}
\date{December 1993}
\maketitle

\begin{abstract}
 I rebuild a conventional two-Higgs doublet model by relaxing the spontaneous
CP violation  and considering approximate global U(1) family symmetries.
So that the domain-wall problem does not explicitly arise at the weak scale,
but CP violation still solely originates from a single CP-phase in the vacuum
after spontaneous symmetry breaking. With this phase four types of
CP-violating mechanism are induced in the model. In particular, by a new
type of the mechanism, both the indirect- and direct- CP violation (i.e.
$\epsilon$ and $\epsilon'/\epsilon$) in kaon decay and  the neutron electric
dipole moment can be consistently accommodated. The masses of the exotic
scalars are weakly constrained in the model and searching
for these particles is worthwhile in the presently accessible energy range.
Substantial CP violation  may occur in the heavy quark and lepton sectors and
probing their effects provides a challenge at B-factory and colliders.
\end{abstract}

\narrowtext

One of the simplest extentions of the standard $SU(2)_{L}\times U(1)_{Y}$
model \cite{WSG} is the conventional two-Higgs doublet model (2HDM).
Recently,  I investigated \cite{WU1} one of the simplest cases in the 2HDM,
that is,  CP is broken spontaneously \cite{TDL} and the
neutral currents conserve all flavors \cite{GW,EAP} at tree level.
Consequently such a simple 2HDM, in which Cabibbo-Kobayashi-Maskawa (CKM)
matrix \cite{CKM} is known to be real\cite{BRANCO}, possesses a new type of
CP-violating mechanism, by which the CP-violating parameters
$\epsilon$ \cite{CCFT} and $\epsilon'/\epsilon$ \cite{RW}
in kaon decay and the neutral electric dipole moment\cite{SMITH} can be
consistently accommodated. In particular, the constraint
on the masses of the exotic scalars is very weak in that model.

   These observations are, on the one hand,  of interest and fascinating in
physical phenomena, on the other hand they bring forth the necessity to
further justify and improve that model. As it is known that there are two
essential issues which need to be clarified. They are the domain-wall problem
\cite{KOZ} and radiative stability\cite{GKEGN}.

    Naturally, a simple and interesting way is to further relax
the conditions of the spontaneous CP violation (SCPV) and the neutral flavor
conservation (NFC) in such a consideration that  the improved model remains
possessing its initial attractive features in the physical phenomena,
and having the basic requirement that CP violation solely originates from the
vacuum, namely if vacuum has no CP violation then the theory becomes CP
invariant (but CP is not necessary to be broken spontaneously). Indeed, the
answer is positive. I shall describe in detail this consideration below.

  It is clear that in the limit that CKM matrix is unity, the conventional 2HDM
with NFC at tree level generates global U(1) family symmetries and the
stability becomes manifest. In the realistic case, it is known that CKM
matrix deviates only slightly from unity.  This implies
that at the electroweak scale any successful models
can only possess approximate global U(1) family symmetries (AGUFS).

    In general, without  imposing any additional conditions, the AGUFS should
also play a role on the neutral currents. Therefore it is natural to assume,
instead of demanding NFC, a partial conservation of neutral flavor (PCNF).
Clearly, the radiative stability now becomes manifest due to the existence
of the small terms of the flavor-changing neutral scalar interactions
(FCNSI)  at tree level. The smallness of the CKM mixings and FCNSI can be
regarded as being naturally in the sense of 't Hooft's criterion\cite{THOOFT}.
It should be mentioned that  the motivation that imposing approximate global
symmetries on the Yukawa couplings is not novel and has been in fact
discussed in the literature. The examples are the approximate flavor
symmetries \cite{HALL} and the approximate discrete symmetries \cite{LW}.
Nevertheless, different considerations will result in different
parameterizations and structures of the Yukawa coupling matrices,
consequently the resulting physical phenomena should be also distinguished.

   Motivated from  the general condition\cite{WU1} of NFC at tree level,
I shall present in this paper an alternative parameterization for
the Yukawa coupling matrices by considering the AGUFS and PCNF. It will be
seen that this type of parameterization provides a more conventional and
general structure for the Yukawa coupling matrices, which will be found to be
very useful in classifying various CP violations
and  analysing possible new physical phenomena. In particular, the
suppressions of the FCNSI become more manifest, as a consequence, the
constraint on the masses of the exotic scalars can become very weak in this
model, so that the masses of these exotic scalars can be just around the
present experimental bound.

 To prevent the domain-wall problem from arising explicitly in the model,
I come to the following observation that

   {\it In the gauge theories of spontaneous symmetry breaking (SSB),
CP violation can be required solely originating from the vacuum after SSB,
even if CP symmetry is not good prior to the symmetry breaking.} In other
words, there exists a kind of explicit CP violation which can be attributed
to the one in the vacuum after SSB, namely if vacuum conserves CP then such an
explicit CP violation  disappears simultaneously and the theory becomes
CP invariant.

     In general, the {\it demanded condition} for such a statement is:
{\it CP nonconservation occurs only at one place of the interactions in the
Higgs potential}.  In particular, this condition may be simply realized by an
{\it universal rule} that in a renormalizable lagrangian all the  interactions
with dimension-four  conserve CP and only  interactions with dimension-two
possess CP nonconservation. It may also be naturally implemented through
imposing some symmetries. An interesting example  is the
Weinberg 3HDM\cite{SW1} in which one can always choose a basis so that there
is only one place allowing to violate CP because of discrete symmetries.
To distinguish from the SCPV, such a CP violation may be refered as a
Vacuum CP Violation (VCPV).

 As an interesting case, let me directly write down in the 2HDM the most
general Higgs potential subject to gauge invariance and the universal rule
stated above.

\begin{eqnarray}
V(\phi) & = & \lambda_{1}(\phi_{1}^{\dagger}\phi_{1}- \frac{1}{2} v_{1}^{2})^2
+ \lambda_{2}(\phi_{2}^{\dagger}\phi_{2} - \frac{1}{2} v_{1}^{2})^2 \nonumber
\\
& & + \lambda_{3}(\phi_{1}^{\dagger}\phi_{1} - \frac{1}{2} v_{1}^{2})
(\phi_{2}^{\dagger}\phi_{2} - \frac{1}{2} v_{2}^{2})
+ \lambda_{4}[(\phi_{1}^{\dagger}\phi_{1})(\phi_{2}^{\dagger}\phi_{2})
-(\phi_{1}^{\dagger}\phi_{2})(\phi_{2}^{\dagger}\phi_{1})]  \nonumber  \\
& &  + \frac{1}{2}\lambda_{5}(\phi_{1}^{\dagger}\phi_{2} +
\phi_{2}^{\dagger}\phi_{1} - Re (\hat{v}_{1}^{\ast} \hat{v}_2) )^2
+ \lambda_{6}(\phi_{1}^{\dagger}\phi_{2}- \frac{1}{2} (\hat{v}_{1}^{\ast}
\hat{v}_2) ) (\phi_{2}^{\dagger}\phi_{1} - \frac{1}{2} (\hat{v}_{2}^{\ast}
\hat{v}_1) )    \\
& & + [\lambda_{7}(\phi_{1}^{\dagger}\phi_{1} - \frac{1}{2} v_{1}^{2})
+ \lambda_{8} (\phi_{2}^{\dagger}\phi_{2} - \frac{1}{2} v_{2}^{2})]
[\phi_{1}^{\dagger}\phi_{2} + \phi_{2}^{\dagger}\phi_{1} -
Re (\hat{v}_{1}^{\ast} \hat{v}_2)]  \nonumber
\end{eqnarray}
where the $\lambda_i$ ($i=1, \cdots, 8$) are all real parameters.
If all the $\lambda_i$ are non-negative the minimum
of the potential then occurs at $<\phi_{i}^{0} > = \hat{v}_i /\sqrt{2}
\equiv v_i e^{i\delta_i}/\sqrt{2} $ with choosing $\delta_{1}=\delta $ and
$\delta_2 =0$.  It is clear that in the above potential CP nonconservation can
only occur through the vacuum, namely $Im (\hat{v}_{1}^{\ast} \hat{v}_{2})
\neq 0$. Obviously, such a CP violation appears as an explicit one in
the potential provided $\lambda_6 \neq 0$, therefore the domain-wall
problem does not explicitly arise. Nevertheless, a further quantitative
calculation should be of interest. A similar potential with $\lambda_{7} =
\lambda_{8} = 0$ due to discrete symmetrty was also considered in\cite{HG}
for strong CP, where discrete symmetry is  softly
broken by dimension-two terms, its relevant CP violation was
refered as a soft CP violation (SOCPV) which is one of the special cases of
VCPV and makes sense only for that case.  CP-violating effects in our case
are no longer soft due to the hard violating terms of discrete
symmetry in the Yukawa interactions (in fact no discrete symmtry is imposed).
Therefore it is better to refer it as VCPV.

    Let me now present a detailed description for the model with
VCPV and PCNF by starting with the general Yukawa interactions
\begin{equation}
L_{Y} = \bar{q}_{L}\Gamma^{a}_{D} D_{R}\phi_{a} + \bar{q}_{L}\Gamma^{a}_{U}
U_{R}\bar{\phi}_{a} + \bar{l}_{L}\Gamma^{a}_{E} E_{R}\phi_{a} + H.C.
\end{equation}
where $q^{i}$, $l^{i}$ and $\phi_{a}$ are $SU(2)_{L}$ doublet quarks,
leptons and Higgs bosons, while $U^{i}_{R}$, $D^{i}_{R}$ and $E^{i}_{R}$
are $SU(2)_{L}$ singlets.  $\Gamma^{a}_{F}$ ($F= U, D, E$) are the arbitray
real Yukawa coupling matrices. $i = 1,\cdots , n_{F}$ is a family label and
$a = 1, \cdots , n_{H}$ is a Higgs doublet label.

    I now come to make another essential step, namely to parameterize
$\Gamma_{F}^{a}$ in such a conventional way that the global U(1) family
symmetry violations in the charged currents
and the neutral currents  can be easily distinguished and characterized by the
different sets of parameters. Motivated from the general condition \cite{WU1}
of NFC at tree level, it is evident to make the following parameterization
for $\Gamma_{F}^{a}$

\begin{eqnarray}
\Gamma^{a}_{F} & = & O_{L}^{F} \sum_{i,j=1}^{n_{F}}\{ \omega_{i} (g_{a}^{F_{i}}
\delta_{ij}  + \zeta_{F} \sqrt{g^{F_{i}}} S_{a}^{F}\sqrt{g^{F_{i}}} )
\omega_{j} \} (O_{R}^{F})^{T} \\
 & & g^{F_{i}} = | \sum_{a} g^{F_{i}}_{a} \hat{v}_{a} |
/ (\sum_{a} |\hat{v}_{a}|^{2})^{\frac{1}{2}} \nonumber
\end{eqnarray}
with $\{\omega_{i}, i=1, \cdots,n_{F}\}$ the set of diagonalized
projection matrices $(\omega_{i})_{jj'} = \delta_{ji}\delta_{j'i}$.
$\hat{v}_{a}$ ($a=1, \cdots , n_{H}$) are the VEV's which will develop from
the Higgs bosons after SSB. $g^{F_{i}}_{a}$ are the
arbitrary real Yukawa coupling constants. $S_{a}^{F}=0$ for $a=n_{H}$ and
$S_{a}^{F}$ ($a\neq n_{H}$) are the arbitrary off-diagonal real matrices.
$g^{F_{i}}$ are  introduced so that a comparison between the diagonal and
off-diagonal matrix elements becomes  available. $\zeta_{F}$ is a conventional
parameter introduced to scale the off-diagonal matrix elements with
$(S_{1}^{F})^{12} \equiv 1$ and others $(S_{a}^{F})_{ij}$ being expected to
be of order unity (of course, some elements of $S_{a}^{F}$ may be off by a
factor of 2 or more).  $O_{L,R}^{F}$ are the arbitrary orthogonal matrices.
Generally, one can choose, by a redifinition of the  fermions, a basis so
that $O_{L}^{F}=O_{R}^{F}\equiv O^{F}$ and $O^{U} = 1$ (or $O^{D} = 1$) .
The AGUFS and PCNF then imply that
\begin{equation}
(O^{F})_{ij}^{2} \ll 1 \  , \qquad i\neq j \ ; \qquad  \zeta_{F}^{2}  \ll 1 \
{}.
\end{equation}
where $O^{F}$ describe the AGUFS in the charged currents and $\zeta_{F}$
mainly characterizes the PCNF. Obviously, if taking $\zeta_{F} =0$ , it turns
to the case  of NFC at tree level. When $\zeta_{F} =0$ and $O^{U} = O^{D} = 1$,
the lagrangian then possesses global $U(1)$ family symmetries, i.e.,
$(U, D)_{i} \rightarrow e^{\alpha_{i}} (U, D)_{i}$.

     For the simplest 2HDM, the physical basis  after SSB is defined through
$f_L = (O_{L}^{F}V_{L}^{f})^{\dagger}F_L$ and $f_R = (O_{R}^{F}P^f
V_{R}^{f})^{\dagger}F_R$ with $V_{L,R}^{f}$ being unitary matrices and
introduced to diagonalize the mass matrices
\begin{equation}
(V_{L}^{f})^{\dagger}(\sum_{i}m_{f_{i}}^{o}\omega_{i} + \zeta_{F} c_{\beta}
\sum_{i,j} \sqrt{m_{f_{i}}^{o}} \omega_{i} S_{1}^{F} \omega_{j}
\sqrt{m_{f_{j}}^{o}} e^{i\sigma_{f}(\delta - \delta_{f_{j}})}) V_{R}^{f} =
\sum_{i} m_{f_{i}}\omega_{i}
\end{equation}
with $m_{f_{i}}$ the masses of the physical states $f_{i}= u_i, d_i, e_i$.
Where the following definitions are introduced
\begin{equation}
(c_{\beta} g_{1}^{F_{i}}e^{i\sigma_{f}\delta} +
s_{\beta} g_{2}^{F_{i}})v \equiv \sqrt{2}m_{f_{i}}^{o} e^{i\sigma_{f}
\delta_{f_{i}}}
\end{equation}
with $v^2 = v_{1}^{2} + v_{2}^{2}= (\sqrt{2}G_{F})^{-1}$,
$g^{F_{i}} v = \sqrt{2}m_{f_{i}}^{o}$ and
$ c_{\beta}\equiv \cos\beta = v_1/v$ and $s_{\beta}\equiv \sin\beta
= v_2/v$. Where $P^{f}_{ij} = e^{i\sigma_{f} \delta_{f_{i}}}\delta_{ij}$,
with $\sigma_{f} =+$, for $f= d, e$, and  $\sigma_{f} = - $, for $f = u$.

As a convention, writing $V_{L,R}^{f} \equiv  1 + \zeta_{F} T_{L,R}^{f}$,
 thus the scalar interactions of the fermions can be
written in the physical basis into
\begin{equation}
L_{Y} \equiv  L_{Y}^{o} +  L'_{Y} (\zeta_{F})
\end{equation}
where $L_{Y}^{o}$ possesses no FCNSI and has a simple structure
\begin{eqnarray}
L_{Y}^{o} & = & (2\sqrt{2}G_{F})^{1/2}\sum_{i,j}^{3}\{ \xi_{d_{j}}
\bar{u}_{L}^{i} V_{ij}^{o} m_{d_{j}}d^{j}_{R}H^+
- \xi_{u_{j}}\bar{d}_{L}^{i} V_{ij}^{o T} m_{u_{j}}u^{j}_{R}H^- \nonumber \\
& & + \xi_{e_{j}}\bar{\nu}_{L}^{i} \delta_{ij} m_{e_{j}}e^{j}_{R}H^+ +  H.C. \}
+ (\sqrt{2}G_{F})^{1/2}\sum_{i}^{3} \sum_{k}^{3} \{ \eta_{u_{i}}^{(k)}
m_{u_{i}} \bar{u}^{i}_{L} u^{i}_{R}  \\
& & +  \eta_{d_{i}}^{(k)}m_{d_{i}} \bar{d}^{i}_{L} d^{i}_{R} +
\eta_{e_{i}}^{(k)}m_{e_{i}} \bar{e}^{i}_{L} e^{i}_{R} + H.C. \}H_{k}^{0}
 \nonumber
\end{eqnarray}
and
\begin{equation}
 \xi_{f_{i}} =  \frac{\sin\delta_{f_{i}}}
{s_{\beta}^{2}\sin\delta}e^{i \sigma_{f}(\delta - \delta_{f_{i}})}
\tan\beta - \cot\beta  \  ; \qquad  \eta_{f_{i}}^{(k)} = O_{2k}^{H} +
(O_{1k}^{H} + i \sigma_{f} O_{3k}^{H} ) \xi_{f_{i}}
\end{equation}
where $V^{o} = (O^{U}_{L})^{T}O^{D}_{L}$ is real.   $H^{0}_{k}=(h, H, A)$ are
the three physical neutral scalars and $O_{kl}^{H}$ is the $3\times 3$
orthogonal mixing matrix among these three scalars, $H$ plays the role of
the Higgs boson in the standard model.

$L'_{Y}$ contains FCNSI, its complete experession and physical phenomena will
 be presented in a longer paper\cite{YLWU3}. I mention here only their main
features. Firstly, it is in favor of having $\tan\beta > 1$ and
$\sin\delta_{f_{i}}/\sin\delta \alt 1$ in order to have the FCNSI be suppressed
manifestly. To the first order in $\zeta_{F} $ the couplings of the FCNSI
are of order O($\zeta_{F} s_{\beta}^{-1}(\sqrt{2}G_{F}m_{f_{i}}
m_{f_{j}})^{1/2}
(S_{1}^{F})_{ij}$).  When $\zeta_{D} \alt 10 ^{-3} m_{H^{0}}(GeV)/(1GeV)$ the
masses of the exotic scalars are unconstrained from the $K^{0} - \bar{K}^{0}$
and $B^{0} - \bar{B}^{0}$ mixings. For a typical value $\zeta_{D} \sim V_{cb}
\sim 0.04$, the masses of the exotic scalars are allowed to be just around the
present experimental bound ($m_{H} \sim 48 $GeV). Secondly, the CP-violating
parameter $\epsilon$  and the mass difference between the $K_{L}$ and $K_{S}$
may be accommodated by FCNSI through fine-tuning the CP phases
$\delta$, $\delta_{s}$, $\delta_{d}$ and the phases $arg (O_{1k}^{H} + i
O_{3k}^{H} )$  arising from the scalar-pseudoscalar mixings (SPM). But its
contribution to $\epsilon'/\epsilon$ is in general small.  Finally, the
smallness of CP violation  becomes natural in the induced KM-type mechanism.
This is  because in the present model CP-phase in the KM-matrix
$V=(V^{u}_{L})^{\dagger} V^{o} V^{d}_{L}$  is  directly related to FCNSI.
In a good approximation, to the first order of
$\zeta_{F} T_{L,R}^{f}$ the amplitudes of the imaginary part in V are of order
O( $\zeta_{F} c_{\beta} (m_{f_{i}}/m_{f_{j}})^{\frac{1}{2}} (S_{1}^{F})_{ij}$)
with $i<j$.

  In general, there appear four types of CP-violating mechanism
which are induced from the single CP-phase $\delta$ in the vacuum. Three of
them (i.e. FCNSI, KM-type and SPM) have been mentioned above. Let me now
concentrate in this short paper on the new type of CP violation mechanism in
$L^{o}_{Y}$ provided $\zeta_{F} < 10^{-3}$ so that the effects from the FCNSI
and KM-type mechanisms are negligible. Without making additional
assumptions, $m_{f_{i}}$, $V_{ij}^{o}$,  $m_{H_{k}^{0}}$, $\xi_{f_{i}}$
(or $\delta_{f_{i}})$, $m_{H^{+}}$ and $O_{kl}^{H}$ (or $\eta_{f_{i}}^{(k)}$)
are in general all the free parameters and will be determined only by the
experiments. An important feature of the present model is that rich
CP-violating
phases ($\delta_{f_{i}}$) are induced from the single CP-phase
$\delta$. They are in general all different and observable.
Clearly, it is  unlike the Weinberg 3HDM\cite{SW1} in
which it is equivalent to the case that $\xi_d = \xi_s = \xi_b$ and
$\xi_u = \xi_c = \xi_t$, therefore the difficulties  encounted \cite{SW2,HYC}
in the Weinberg 3HDM do not arise in the present model.

  Let us first check the CP-violating parameters $\epsilon$ and
$\epsilon'/\epsilon$ in kaon decays. Like the Weinberg 3HDM, but unlike the
KM-model, the long-distance contribution to $\epsilon$ become important.
The imaginary part of the $K^0-\bar{K}^0$ mixing mass matrix to which
$\epsilon$ is propotional mainly receives the contribution from the $\pi$,
$\eta$ and $\eta'$ poles\cite{HAG}. Following the analyses in the
literature\cite{HAG,DH,HYC}, we have in our present model
\begin{equation}
 \epsilon \simeq 2.27 \times 10^{-3} Im(\xi_s\xi_{c}-0.05\xi_{d}^{\ast}
\xi_{c}^{\ast})\frac{37GeV^{2}}{m_{H^{+}}^{2}}(ln\frac{m_{H^{+}}^{2}}
{m_{c}^{2}}-\frac{3}{2})
\end{equation}
which should easily accommodate the experimental value $|\epsilon| = 2.27\times
10^{-3}$ since $\xi_s$ and $\xi_c$ are all free parameters. It will be shown
\cite{YLWU3} that the short-distance box graphs with charged-scalar in our
model can also provide a large contribution to $\epsilon$.

   The ratio $\epsilon'/\epsilon$ in the Higgs-boson-exchange models
 was first correctly estimated by Donoghue and Holstein \cite{DH}. The
evaluations can directly be applied to the present model. In fact,
it was shown in the refs. \cite{DH,HYC} that the ratio does not depend on
the detail of the CP-odd matrix element and is given by
\begin{equation}
\epsilon'/\epsilon = 0.017 D = (0.4-6.0)\times 10^{-3}
\end{equation}
which is comparable to the one calculated from the standard KM-model\cite{WU3}
and is also consistent with the present experimental data\cite{RW}.
 A short-distance contribution to the ratio from charged-scalar exchange at
tree level is also found to be of order $10^{-3}$ \cite{YLWU3}.

     Consider now the neutron EDM, $d_{n}$. The present experimental limit
on $d_{n}$ is $d_{n} < 1.2 \times 10^{-25}$ {\it e cm} \cite{SMITH}.
Applying various well-known scenarios for the calculations of $d_{n}$ to
the present model, the $d_{n}$ is then accommodated by choosing the
parameters $\xi_{f_{i}}$ and $\eta_{f_{i}}^{(k)}$ to satisfy the following
conditions
\begin{eqnarray}
 & & D1: \qquad  Im(\xi_d\xi_{c})\frac{1}{m_{H^{+}}^{2}}
(ln\frac{m_{H^{+}}^{2}}{m_{c}^{2}}-\frac{3}{4})
\alt 6.3\times 10^{-2} GeV^{-2}\ ,   \\
& &  D2: \qquad Im [\eta_{t}^{(k)}]^{2} h_{NH}(m_{t},m_{H_{k}^{0}})
\alt 0.18 \ ,  \\
 & &  D3: \qquad  Im(\xi_b\xi_{t})h_{CH}(m_{t},m_{b},m_{H})
\alt 3.0\times 10^{-2} \ ,  \\
 & & D4: \qquad  Im(\eta_{d}^{(k)} \eta_{t}^{(k)} + 0.5\eta_{u}^{(k)}
\eta_{t}^{(k)}) \alt 0.2 \ , \qquad  (m_{t}\sim m_{H})
\end{eqnarray}
where the condition $D1$ is from the quark model through charged-Higgs boson
exchange,  $D2$ and $D3$ from the Weinberg's  gluonic operator through
the neutral-Higgs boson exchange\cite{SW2} and  the charged-Higgs boson
exchange\cite{DICUS} respectively . $h_{NH}$ and $h_{CH}$ are the functions
of the quark- and Higgs-mass arising from the integral of the loop.
$h_{NH}\simeq 0.05$ for $m_{t}=m_{H}$. $h_{CH}\leq 1/8$ and $h_{CH}=1/12$ for
$m_{b}\ll m_{t}=m_{H}$. $D4$ is from the gluonic
chromoelectric dipole moment (CEDM)\cite{DWC} induced by the Barr-Zee two-loop
mechanism\cite{BZ}.

  For the present experimental bound of $m_{H^{+}}\agt 45$GeV, it is not
difficult to find that $Im\xi_{s}\xi_{c} \agt 10 $ and
$Im\xi_{d}\xi_{c}\sim 21$. A natural solution for them is
$\tan\beta \gg 1$, i.e., $v_{2} \gg v_{1}$, such a hierarchy was
already discussed in\cite{BBG}.

   The hierarchic structure of the VEV's also implies that  large CP violation
may occur in the heavy quark and lepton sectors due to their possible large
complex Yukawa couplings. One of the interesting examples is the T-odd and
CP-odd triple momentum correlations in the exclusive semileptonic B-meson
decay  $B\rightarrow D^{\ast} (D \pi ) \tau \nu_{\tau}$ \cite{WU4}. Using the
analysis of the ref. \cite{WU4} to the present model, the CP asymmetry reads
\begin{equation}
A_{V_{T}P} \simeq 10^{-2} \frac{m_{b}m_{\tau}}{m_{H^{+}}^{2}}
Im (\xi_{b}\xi_{\tau}^{\ast} ) \alt 10^{-2} - 10^{-3}
\end{equation}
where the last numerical values are estimated by taking $|\xi_b| \sim
|\xi_{\tau}| \sim (v_2 /v_1 ) \sim (m_t /m_b )$, $m_{H^{+}}\sim m_{W}$ and
$\sin (\delta_{\tau}-\delta_{b})\sim 1.0 - 0.1$.

Before ending this paper,  I would like to briefly remark that the strong CP
problem may be simply evaded by using the well-known Peccei-Quinn mechanism
\cite{PQ} realized in a heavy fermion invisible axion scheme\cite{KSVZ}.
Since this scheme, on the one hand, is one of the simplest schemes and has an
advantage that it leaves the above model almost unchanged, and on the other
hand it is also free from the axion domain-wall problems.

    In a word, the above considerations indicate that the conventional 2HDM
with VCPV and PCNF may provide one of the simplest and attractive models in
understanding the origin and mechanisms of CP violation. It also opens a
window for  probing new physical phenomena at B-factory and colliders.
In particular, the weak constraint on the masses of the exotic scalars makes
the searching for these particles in the presently accessible energy range
more worthwhile. Various interesting features arising from this model
are going to be discussed in a longer paper\cite{YLWU3}.

   I am grateful to Lincoln Wolfenstein for valuable discussions and for
reading the manuscript. I also wish to thank Ling-Fong Li and Jiang Liu
for useful comments. This work was supported by DOE grant \# DE-FG02-91ER40682.

\end{document}